\documentclass{ws-mpla}

\usepackage[mathlines,pagewise]{lineno}
\usepackage[super,nosort,nospace,compress]{cite}
\usepackage[hyphens]{url}
\usepackage[T1]{fontenc}
\usepackage[breaklinks]{hyperref}

\def\refcite#1{\def\citepunct{,\ }\citen{#1}\def\citepunct{,}}
\def\myfig#1{\centerline{\psfig{file=\ArtDir/#1}}}

\def\boldendash{\leavevmode\lower-.8pt\hbox{\kern-0.1pt	      
	\vrule height 2.3pt depth -1.6pt width 6pt}}	      
\def\ccm{}
\def\afm{}
\def\asm{}
\def\jat{}
\def\org{}
\def\vol{}
\def\efm{}
\def\pgs{}
\def\at{}
\def\y{}
\def\pt{}
\def\cp{}
\def\pp{}
\def\esm{}
\def\pn{}
\def\pti{}

\begin{document}
\sloppy
\edef\logoname{146br}
\def\artid{1330027}
\def\ArtDir{30027}


\setcounter{page}{1}

\markboth{S. L. Wu}
{Brief History for the Search and the Discovery of the Higgs Particle}



\title{Brief history for the search and discovery\\ of the Higgs particle -- A personal perspective}

\author{Sau Lan Wu}

\address{Department of Physics, University of Wisconsin, Madison, WI 53706, USA\\
\email{sau.lan.wu@cern.ch}}

\maketitle

\begin{history}
\received{}
\accepted{}
\end{history}

\begin{abstract}
In 1964, a new particle was proposed by several groups to answer the question of where the
masses of elementary particles come from; this particle is usually
referred to as the Higgs particle or the Higgs boson. In July
2012, this Higgs particle was finally found experimentally, a feat
accomplished by the ATLAS Collaboration and the CMS Collaboration using
the Large Hadron Collider at CERN.  It is the purpose of this review to give my personal perspective on a brief history of the experimental search for this particle
since the '80s and finally its discovery in 2012.  Besides
the early searches, those at the LEP collider at CERN, the Tevatron
Collider at Fermilab, and the Large Hadron Collider at CERN are described
in some detail.  This experimental discovery of the Higgs boson is often
considered to be the most important advance in particle physics in the
last half a century, and some of the possible implications are briefly
discussed. 
This review is partially based on a talk presented by the author at the conference ``Higgs Quo Vadis,'' Aspen Center for Physics, Aspen, CO, USA, March 10-15, 2013.
\end{abstract}


\ccode{PACS Nos.: 14.80.Bn, 13.85.-t, 13.66.Fg, 12.60.-i}

\section{Introduction}\label{sec:Introduction}
\subsection{Fundamental interaction and gauge particles}\label{sec1.1}
There are four types of interactions in nature: gravitational, weak, electromagnetic and strong.  In the study of elementary particles at high energy, the gravitational interaction plays no known role. The other three interactions all proceed through the exchange of gauge particles.

The gauge particle for the electromagnetic interaction is the best-known one: the photon, as first pointed out in one of the famous 1905 papers of Einstein\cite{Einstein}.  While the photon does not interact with itself, all the other gauge particles have self-interactions, i.e.~they are all Yang--Mills particles\cite{Yang}.

The gauge particle for strong interactions --- the gluon --- is the second gauge particle, and the first Yang--Mills particle, to be observed experimentally and\break directly\cite{Wu,Brandelik,PIEC}.  Four years later, the gauge particles $W$ and $Z$ for weak interactions were observed at CERN\cite{Arnisson,Banner}. Some properties of these gauge particles are listed in Table~\ref{tab:properties}. All gauge particles have spin~1.

\begin{table}[t]
\tbl{Some of the observed properties of the gauge particles\protect\cite{phys} --- all have spin~1 and baryon number~0.\label{tab:properties}}
{\tabcolsep23pt
\begin{tabular}{@{}lcc@{}l}
\Hline\\[-6pt]
Particle &Electric charge $(e)$ &Mass (GeV/$c^2$)\\[3pt]
\hline\\[-6pt]
$g$ (gluon) &0 &0 \\[5pt]
$\gamma$ (photon) &0 &0 \\[5pt]
$W^\pm$  &$\pm1$ &$80.385\pm 0.015$ \\[5pt]
$Z$ &0 &$91.1876\pm 0.0021$ \\[3pt]
\Hline
\end{tabular}}
\end{table}

\begin{table}[b]
\tbl{Some of the observed properties of the quarks and leptons\protect\cite{phys} --- all have spin $\frac12$. (Antiquarks and antileptons have the opposite charges and baryon numbers.)\label{tab:quarks}}
{\tabcolsep5.5pt
\begin{tabular}{@{}llccc@{}}
\Hline\\[-6pt]
Generation & Particle &Electric charge $(e)$ &Baryon number &Mass (GeV/$c^2$) \\[3pt]
\hline\\[-6pt]
\multicolumn{1}{@{}c}{\phantom{II}I} & $u$ (up quark) & $+\frac{2}{3}$ & $\frac{1}{3}$ & $0.0023^{+0.0007}_{-0.0005}$ \\[8pt]
& $d$ (down quark) & $-\frac{1}{3}$ & $\frac{1}{3}$ & $0.0048^{+0.0007}_{-0.0003}$ \\[8pt]
& $\nu_{e}$ (electron neutrino) & 0 & 0 & small \\[8pt]
& $e$ (electron) &$-1$ & 0 &$0.000511$ \\[8pt]
\multicolumn{1}{@{}c}{\phantom{I}II}  & $c$ (charm quark) & $+\frac{2}{3}$ & $\frac{1}{3}$ &$1.275\pm 0.025$ \\[8pt]
& $s$ (strange quark) & $-\frac{1}{3}$ & $\frac{1}{3}$ &$0.095\pm 0.005$ \\[8pt]
& $\nu_{\mu}$ (muon neutrino) &0 &0 &small \\[8pt]
& $\mu$ (muon) &$-1$ & 0 &$0.105658$ \\[8pt]
\multicolumn{1}{@{}c}{IIII} & $t$ (top quark) &$+\frac{2}{3}$ & $\frac{1}{3}$ &$173.5\pm 1.0$ \\[8pt]
& $b$ (bottom quark) &$-\frac{1}{3}$ & $\frac{1}{3}$ &$4.65\pm 0.03$ \\[8pt]
& $\nu_{\tau}$ (tau neutrino) & 0 & 0 & small \\[8pt]
& $\tau$ (tau) & $-1$ & 0 &$1.77682\pm 0.00016$ \\[3pt]
\Hline
\end{tabular}}
\end{table}

\subsection{Quarks and leptons}\label{sec1.2}
Atoms and molecules are composed of protons, neutrons and electrons; protons and neutrons are made of the $u$ and the $d$ quarks of Gell-Mann\cite{GellMann} and Zweig\cite{Zweig}.\break The $u$, the $d$, and the electron are all members of the first generation of quarks and leptons. There are three known generations of quarks and leptons; some of their properties are listed in Table~\ref{tab:quarks}.  Thus there are six quarks and six leptons together with their antiparticles.  All these quarks and leptons have spin $\frac{1}{2}$.

The masses of the neutrinos are much smaller than that of the electron, which is the lightest lepton besides the neutrinos.  Experimentally, it is known that at least two of the three neutrinos ($\nu_{e}$, $\nu_{\mu}$, $\nu_{\tau}$) have nonzero mass; it is generally accepted that all three neutrinos have nonzero masses.

\subsection{Standard model}\label{sec1.3}
The interactions of these quarks and leptons are described by the standard model of Glashow, Weinberg and Salam\cite{Glashow}, which is a gauge theory with the group $U(1) \times \hbox{\it SU}(2) \times \hbox{\it SU}(3)$, where the gauge particles are those listed in Table~\ref{tab:properties}.

It is an important feature of the standard model that the left-handed and the right-handed components of the quarks and leptons interact differently, according to the $SU(2)$ and the $U(1)$ parts of the gauge group, respectively.  This has far-reaching consequences.  If a quark or a lepton has nonzero mass (which is indeed the case as seen from Table~\ref{tab:quarks}), then the right-handed and the left-handed components can be transformed into each other through a Lorentz transform.  This important point can be seen more explicitly as follows.  Expressed in term of the left-handed and the right-handed components, the mass term of an electron, for example, is
\begin{eqnarray*}
m_{e}\bar{\psi}_{e}\psi_{e} &=&m_{e} (\bar{\psi}_{eL}+\bar{\psi}_{eR}) (\psi_{eL}+\psi_{eR})\\[5pt]
&=&m_{e}\bar{\psi}_{eL}\psi_{eR}+m_{e}\bar{\psi}_{eR}\psi_{eL}\,.
\end{eqnarray*}

Thus, in order to get a mass term for the electron, it is necessary to couple the left-handed and the right-handed components.  But such a coupling is not allowed because these left-handed and right-handed components transform differently under a gauge transform.

The question is therefore:  how can the quarks and the leptons acquire their masses?

The Higgs particle comes to rescue.

\subsection{Higgs particle}\label{sec:higgsparticle}
The basic idea proposed in 1964 is to introduce a new particle or field that has a nonzero vacuum expectation value.  Some of the early papers on this idea are listed as Ref.~\refcite{Englert}. In the context of the standard model, this new particle is commonly called the Higgs particle.

For many years, this Higgs particle had remained the only particle in the standard model not observed experimentally.  It was finally discovered in the summer of 2012 by the ATLAS Collaboration\cite{atlasdisc} and the CMS Collaboration\cite{cmsdisc} at CERN using the proton--proton accelerator called the Large Hadron Collider.  It is the purpose of this review to give a brief history of the search and the discovery of this Higgs particle.

What are the basic properties of this particle?  The most fundamental one is that the Higgs particle, since it has a nonzero vacuum expectation value, must have spin zero.

\baselineskip=12.9pt

As already noted above, all the quarks and the leptons are spin-$\frac{1}{2}$ particles, while all the gauge particles have spin one.  Therefore, this newly discovered Higgs particle not only completes the list of particles in the standard model, but is also the first elementary particle of spin zero.  If one takes the view that this new class of spin-0 elementary particles consists not just of this Higgs particle, then it is possible that this experimental discovery of the Higgs particle may lead to an entirely new era of high-energy physics.

\section{Early Searches for the Higgs Particle}\label{sec:earlysearch}
\subsection{Mass of the Higgs particle}\label{sec2.1}
In particle physics, there is very little understanding of the masses of the various elementary particles.  For example, even though the masses of the electron and the muon have both been measured to the accuracy of eight digits, there is no understanding why the muon is about 200 times heavier than the electron. As another example, the mass of the top quark was not known even roughly until it was observed in Fermilab\cite{Fermilab}.  It is therefore not surprising that there was no guess on the mass of the Higgs particle even many years after it was proposed theoretically.\cite{Englert}

For this reason, the experimental search for this Higgs particle must cover a very large range of mass.

\subsection{A first {\upshape``}signal\/{\upshape ''(1984)}}\label{sec2.2}
\begin{figure}[t]
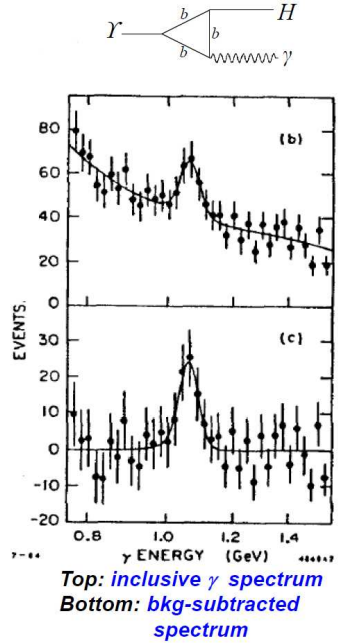

\myfig{30027f001.pdf,width=1.8in}
\caption{Energy spectrum of photons from the $\varUpsilon$ decay, with a fit using a third-order polynomial for the background and the crystal ball line shape for the signal region (top), and the background-subtracted spectrum (bottom).\protect\cite{Crystal} }
\label{fig1}
\end{figure}

Limited by the accelerators available at that time, the early searches, i.e.~the searches carried out in the '80s, are for the case of relatively low masses.  The first excitement came from the Crystal Ball Collaboration using the DORIS electron--positron collider at DESY.  They studied the decay
$$\varUpsilon \rightarrow H + \gamma$$
and found a peak in the $\gamma$ spectrum, corresponding to a Higgs mass of 8.32~GeV/$c^2$.\cite{Crystal} This peak, reported at the 1984 International Conference of High-Energy Physics in Leipzig and shown in Fig.~\ref{fig1}, has a high statistical significance of above $5 \sigma$.

Unfortunately, this exciting result was not confirmed by the CESR at Cornell University. In fact, with more data, the Crystal Ball peak\cite{Crystal} also disappeared.

\subsection{Other searches in the '80s}\label{sec2.3}
None of the other searches in the '80s produced a signal for the Higgs particle.  Here are three examples of such searches.

(a)~The CUSB Collaboration at CESR of Cornell University continued to look for Higgs particle in the upsilon decay $\varUpsilon \rightarrow H + \gamma$, the same channel used by the Crystal Ball Collaboration, but did not find a signal.\cite{CUSB}

\baselineskip=13pt

(b)~The SINDRUM Collaboration at the Paul Scherrer Institute proton cyclotron looked for very low-mass Higgs particle through the decay $\pi^{+}\rightarrow e^{+}\nu_{e}H$ with $H\rightarrow e^{+}e^{-}$, but did not find a signal.\cite{SINDRUM}

(c)~The CLEO Collaboration at CESR of Cornell University searched for the Higgs particle in $B$ decay:
$$B\rightarrow K + H\,,$$
with
$$H\rightarrow \mu^{+}\mu^{-},\quad \pi^{+}\pi^{-}\quad \mbox{or}\ \ K^{+}K^{-}\,,$$
but again did not find any signal.\cite{CLEO}

The conclusion from such searches is that the Higgs mass was likely to be larger than 8~GeV/$c^2$ or 9~GeV/$c^2$.

\section{Search at LEP}\label{sec:SLEP}
\subsection{LEP} \label{sec:LEP}
LEP is the \underline{L}arge \underline{E}lectron \underline{P}ositron collider built at CERN. It was housed in an underground tunnel across the Switzerland--France border; the tunnel has a circumference of 27~km, or 17~miles.

Shortly after taking a leading role in the discovery of the gluon at DESY, I decided to join the ALEPH Collaboration at LEP. My Wisconsin group was the first US group invited to join the ALEPH experiment. The main reason for this move to CERN was my desire to concentrate on discovering experimentally the Higgs particle, but I did not realize at that time this quest was going to take 32~years, from 1980 to 2012!

The LEP design energy for each beam was 100~GeV, meaning that the center-of-mass energy was originally designed to reach 200~GeV. There were four detectors at LEP: ALEPH, DELPHI, L3 and OPAL.

The operation of LEP can be described in terms of two periods, designated as LEP1 and LEP2. The strategies to search for the Higgs particle are quite different, and are to be described in Secs.~\ref{sec:LEP1} and \ref{sec:LEP2}, respectively below.

\subsection{Search at LEP\hspace*{0.5pt}1 {\upshape(1989\boldendash 1995)}} \label{sec:LEP1}
During this period, LEP operated at the $Z$ peak, i.e.~the center-of-mass energy was 91.18~GeV --- see Table~\ref{tab:properties}. Thus the event rate was very high. Furthermore, since the $Z$ mass is much larger than those of the particles used to search for the Higgs particle as described in Sec.~\ref{sec:earlysearch}, LEP1 provided a significantly larger range for the Higgs mass.

\begin{figure}[t]
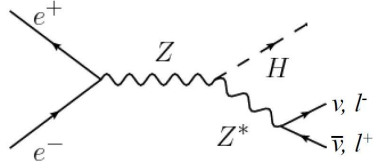

\myfig{30027f002.pdf,width=2.0in}
\caption{The Feynman diagram for the Higgsstrahlung process $e^+e^- \rightarrow Z \rightarrow Z^*H$, with $Z^* \rightarrow \ell^{+}\ell^{-}$ or $Z^* \rightarrow \nu\bar\nu$.}
\label{lep1}
\end{figure}

At this $Z$ peak, the most useful decay channels for the search of the Higgs particle are through
$$Z\rightarrow H \ell^{+}\ell^{-}$$
and
$$Z\rightarrow H\nu\bar\nu\,,$$
where $\ell$ means as usual either the electron or the muon, as shown in Fig.~\ref{lep1}. In the first channel  $Z\rightarrow H \ell^{+}\ell^{-}$, the observation of the two charged leptons in the decay product gives very clean events. The second channel $Z\rightarrow H\nu\bar{\nu}$, on the other hand, has the important advantage of a larger branching ratio.

Plans to search for the Higgs particle at LEP1 were under way in the early '80s. For example, in the 1983 meeting of the LEP committee, the following question --- Question~6 --- was raised:

``What strategy with respect to data acquisition and analysis would you follow to search for Higgs in $Z$ decay? Suppose one needs $10^7$ $Z$ decays to observe ten~events of the type $Z\rightarrow e^{+}e^{-}H$, $H\rightarrow \mbox{hadrons}$, $m_{H}=50~{\rm GeV}/c^2$.''

This question was answered by my Wisconsin group on December~21, 1983\,\cite{Mermikides}{:}

``Starting with $10^7$ $Z^0$ events and by matching the TPC hits in the pad rows to the electromagnetic showers etc., we can reduce the $10^7$ $Z^0$ events to $3\times 10^4$ events with 80\% efficiency for the Higgs events at Higgs mass of 50~GeV/$c^2\ldots$ Full reconstruction of these $3\times 10^4$ events for physics is expected to be 20--30~sec CPU of IBM 370/168 per event.''

Reference~\refcite{LEP} gives a partial list of the publications for the search for the Higgs particle on the basis of the data from LEP1. Due to the difficulties of combining the searches from four experiments, very roughly the data from LEP1 excludes the mass of the Higgs particle to be below 65~GeV/$c^2$.

\subsection{Search at LEP\hspace*{0.5pt}2 {\upshape(1995\boldendash 2000)}} \label{sec:LEP2}
\begin{figure}[t]
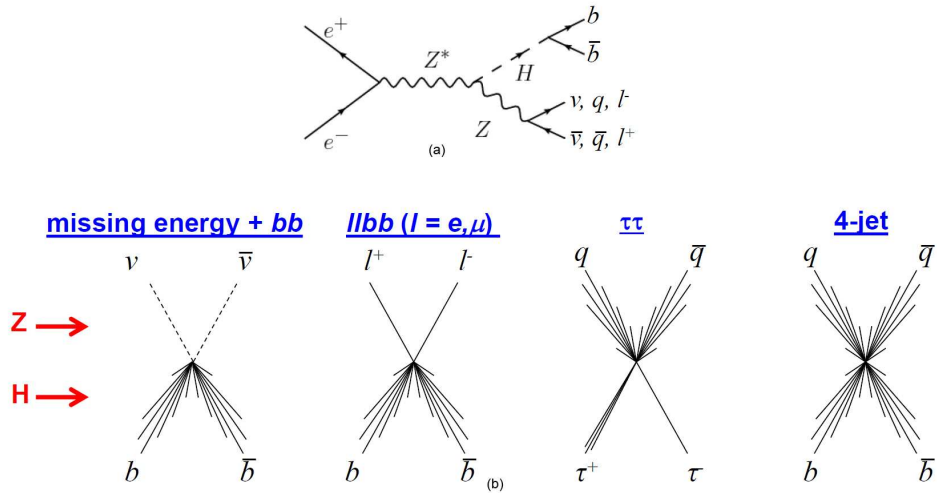

\myfig{30027f003.pdf,width=5.0in}
\caption{Top: the Feynman diagram for the process $e^+e^- \rightarrow Z^* \rightarrow ZH$, with $H \rightarrow b\bar{b}$ and $Z \rightarrow \ell^{+}\ell^{-}$, $Z \rightarrow \nu\bar\nu$ or $Z \rightarrow q\bar{q}$. Bottom: the four major final states for Higgs search at LEP2. The particles from the $Z$ decay and from the Higgs decay are shown. (Figure taken from M. M. Kado and C. G. Tully, Annu. Rev. Nucl. Part. Sci. 2002. 52:65-113)}
\label{fig2}
\end{figure}

For LEP2, the center-of-mass energy went beyond the $Z$ peak, much beyond. Because of this increase in energy, the search at LEP2 differs from that at LEP1 in that the $Z$ is real instead of virtual. More precisely, the search at LEP2 can be represented schematically as that of Fig.~\ref{fig2}. Of these channels, the last one, called the four-jet channel, is the most sensitive one.

\begin{figure}[t]
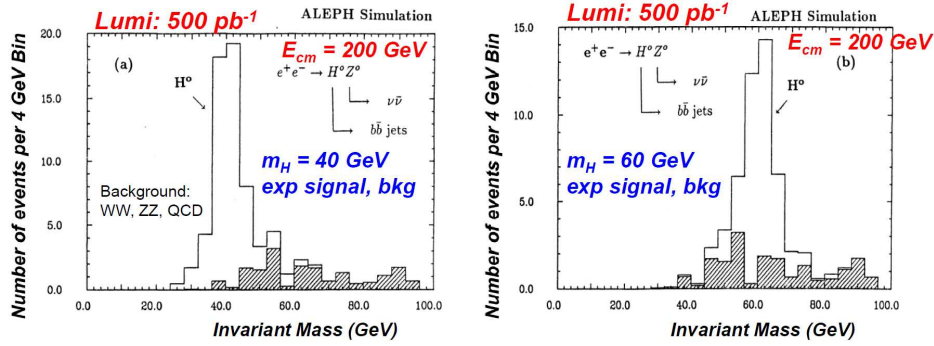

\myfig{30027f004.pdf,width=4.9in}
\caption{Results from simulation showing the invariant mass distribution of the $b\bar{b}$ pair from Higgs decay in the ALEPH experiment at LEP2. The results correspond to an integrated luminosity of 500~pb$^{-1}$ at a center-of-mass energy of 200~GeV. Signal distributions are shown for $m_H=40$~GeV (left) and
$m_H=60$~GeV (right). In both plots, the background distribution is shown as a shaded area.}
\label{fig3}
\end{figure}

In the ECFA workshop of 1986, I presented the simulated results shown in Fig.~\ref{fig3}, where the center-of-mass energy of LEP was taken to be 200~GeV with the Higgs mass of 40~GeV/$c^2$ and 60~GeV/$c^2$.\cite{Sau Lan Wu}

\baselineskip=12.9pt

Until the year 1999, with LEP running at a center-of-mass energy up to 200~GeV, no indication was found of the production of the Higgs particle. The 95\% confidence limit for the lower bound of the Higgs mass was 107.9~GeV/$c^2$. This lower bound is essentially the difference of the LEP energy and the $Z$ mass: $200-91.18=108.82$.

Then things became more interesting.

Gigantic efforts were made to push the LEP energy beyond the original design energy. With new ideas of working with LEP, six machine upgrades were implemented. In this way, the center-of-mass energy reached a phenomenal value of 209~GeV, making LEP sensitive in principle to a Higgs mass of $107.9+9=116.9~{\rm GeV}/c^2$.

Promising candidates started to show up! In late June 2000, ALEPH found the first Higgs candidate with a reconstructed mass of 114~GeV/$c^2$. This event was found by both cut-based and neural network analyses. The ten most significant Higgs candidates from the four experiments at LEP are listed in Table~\ref{tab:candidates}.\cite{ALEPH,Namura}

\begin{table}[t]
\tbl{Some properties of the ten most significant Higgs candidates from LEP2.\protect\cite{ALEPH,Namura}\label{tab:candidates}}
{\tabcolsep8.3pt
\begin{tabular}{@{}clcccc@{}}
\Hline\\[-6pt]
&&&&&$\ln(1+s/b)$\\
&\multicolumn{1}{c}{Expt.} &$E_{\rm cm}$ (GeV) & Decay channel
&$M^{\rm rec}_{H}$ (GeV/$c^2$) &115~GeV/$c^2$ \\[3pt]
\hline\\[-6pt]
\phantom{0}1 & ALEPH   & 206.7 & 4-jet   & 114.3 & 1.73\\[5pt]
\phantom{0}2 & ALEPH   & 206.7 & 4-jet   & 112.9 & 1.21\\[5pt]
\phantom{0}3 & ALEPH   & 206.6 & 4-jet   & 110.0 & 0.64 \\[5pt]
\phantom{0}4 & L3 & 206.4 & E-miss  & 115.0 & 0.53 \\[5pt]
\phantom{0}5 & OPAL    & 206.6 & 4-jet   & 110.7 & 0.53\\[5pt]
\phantom{0}6 & DELPHI  & 206.7 & 4-jet   & 114.3 & 0.49\\[5pt]
\phantom{0}7 & ALEPH   & 205.0 & Lept    & 118.1 & 0.47\\[5pt]
\phantom{0}8 & ALEPH   & 208.1 & Tau     & 115.4  & 0.41\\[5pt]
\phantom{0}9 & ALEPH   & 206.5 & 4-jet   & 114.5 & 0.40\\[5pt]
10 & OPAL   & 205.4 & 4-jet   & 112.6 & 0.40\\[3pt]
\Hline
\end{tabular}}
\end{table}

Because of these events, the LEP Higgs working group made, on November~3, 2000, a request to the LEP committee for a 4--6 months extension in 2001. This request was denied and LEP was shut down in the first week of November 2000, after delivering 11~years of great physics. This is very disappointing for those of us who believed that we were really close to a major discovery.

But, in retrospect, no complaints.

\section{Search at Tevatron}\label{sec4}
\subsection{Tevatron Collider}\label{sec4.1}
The Tevatron Collider was a proton--antiproton colliding accelerator that had been operating at Fermilab since 1987. Its maximum beam energy was 0.98~TeV, and thus its maximum center-of-mass energy was 1.96~TeV. The radius of the tunnel for this collider was 1~km.

\baselineskip=13pt

The advantage of this Tevatron Collider was that it was a proton--antiproton collider. Being a
proton--antiproton collider, both the protons and the antiprotons were accelerated and stored in the same beam pipe.  Thus the Tevatron Collider was the first colliding accelerator in the TeV regime. Since it was a proton--antiproton colliding-beam accelerator, its luminosity is typically lower by a factor of ten compared with that of a proton--proton collider.

\subsection{Detectors and results}\label{sec4.2}
There were two detectors at this Tevatron Collider: CDF and D0.  After LEP was shut down near the end of the year 2000 as described in the preceding section, for ten~years the Tevatron Collider was the only place to search for the Higgs particle. See Sec.~\ref{sec:LargeHadron} below. Unfortunately, during these ten~years, this Tevatron Collider did not manage to discover the Higgs particle; the main reason for this was that it did not produce sufficient amount of integrated luminosity.  It has been estimated that, for the discovery of the Higgs particle, the necessary integrated luminosity is about two and half times of what it actually produced.

Still, the Tevatron experiments produced several results from their Higgs searches; the following is selected from their combined CDF+D0 results:

In July 2011, with up to 8.6~f\hspace*{0.06em}b$^{-1}$, the combined Tevatron analysis was able to exclude (at the 95\%~CL) standard model Higgs boson masses between 156~GeV/$c^2$ and 177~GeV/$c^2$. The result was presented in EPS-HEP 2011.

\begin{figure}[t]
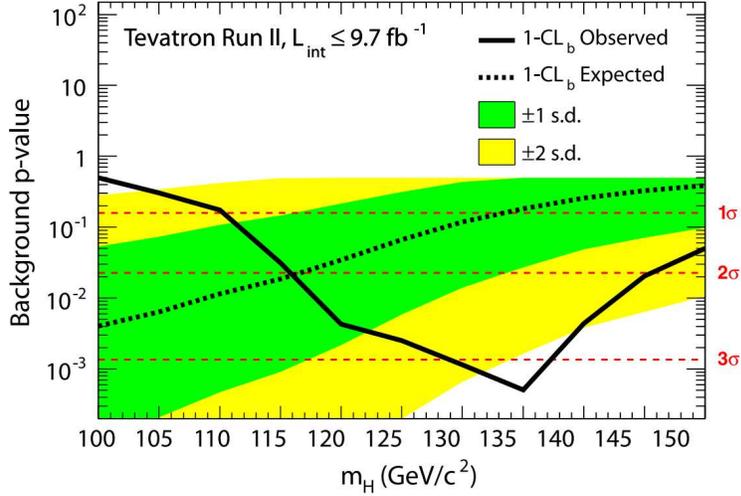

\myfig{30027f005.pdf,width=4in}
\caption{The $p$ value as a function of $m_{H}$ under the background-only hypothesis. Also shown are the median expected values assuming a SM signal is present, evaluated separately at each $m_{H}$. The associated dark- and light-shaded bands indicate the 1~s.d. and 2~s.d. fluctuations of possible experimental outcomes.\cite{CDFD04} }
\label{fig4_1}
\end{figure}

In ICHEP 2012, the Tevatron experiments gave the result of the Higgs search combination with their full dataset (up to 10~f\hspace*{0.06em}b$^{-1}$). They set a 95\%~CL exclusion for Higgs boson masses between 100~GeV/$c^2$ and 103~GeV/$c^2$, and between 147~GeV/$c^2$ and 180~GeV/$c^2$.\cite{CDFD03} More interestingly, they observed a $3\sigma$ excess between 115~GeV/$c^2$ and 140~GeV/$c^2$.\cite{CDFD03} The excess was concentrated in the $H\rightarrow b\bar{b}$\break channel (Fig.~\ref{fig4_1}).\cite{CDFD04}

The Tevatron Collider was shut down in 2011 after 24~years of operation.

\section{Search at the Large Hadron Collider}\label{sec:LargeHadron}
\subsection{Large Hadron Collider}\label{sec:lhc}
The Large Hadron Collider (LHC) is a proton--proton collider built in the LEP tunnel.  Its design energy is 14~TeV in the center-of-mass system; when it began to produce experimental data in 2010, it was running at half of this design energy.  In 2012, this center-of-mass energy was raised to 8~TeV.

This LHC is a remarkable accelerator: since the previous highest energy of 1.96~TeV was from the Tevatron Collider of Fermilab, the initial energy of the LHC was an increase of more than 3.5 times.  Such an increase is phenomenal, and thus there is expectation of a great deal of new physics from LHC.

There are four major experiments at LHC:  two large ones --- ATLAS and CMS~--- together with two smaller ones --- ALICE and LHCb.  The Higgs search and discovery have been carried out entirely by the two large experiments ATLAS and CMS.

In both of these two experiments, the mass range of the Higgs particle covered is from 600~GeV/$c^2$ down to about 110~GeV/$c^2$, which is the lower limit of the Higgs mass from LEP.  Upper Higgs mass limit of about 158~GeV/$c^{2}$ has been known from electroweak fits, but such limit is not taken into account in these searches for two reasons:

\begin{arabiclist}[(2)]
\item[(1)] such a limit is obtained on the assumption that there is only one Higgs particle, but there is no good basis for such an assumption; and
\item[(2)] with the abundance of data from LHC, there is no reason not to carry out the search for the range quoted above.
\end{arabiclist}

\subsection{A major setback}\label{sec:setback}
At the LHC, protons first went around the full ring on September~10, 2008.  Unfortunately, nine days later on September 19, an electric fault triggered a major setback.  The cause of this setback was due to a faulty electric connection between two of the accelerator magnets; this resulted in mechanical damage and the release of a large amount of liquid helium.

A Harvard graduate student on shift when this happened sent the following E-mail:

``LHC E-log: Fire alarm in point~3 and point~4.  Fire brigade going there together with RP.  Massive quench in S34.  Helium released in the tunnel.  This had to happen on my shift, of course.  No beam for at least one week.''

\begin{figure}[t]
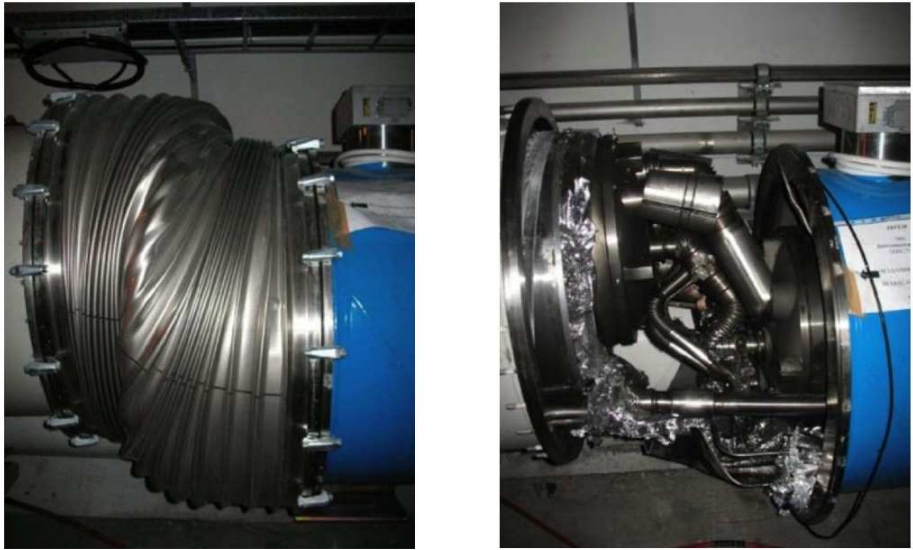

\myfig{30027f006.pdf,width=5.0in}
\caption{The interconnect between two LHC dipole magnets following the incident of September~19, 2008.  The photograph on the left shows one magnet displaced vertically with respect to the other. The sheathing has been removed in  the photograph on the right, in which the damage to the components in the interconnect is visible.} \label{fig4}
\end{figure}

Some of the damages are shown in Fig.~\ref{fig4}.

The damage turned out to be much more extensive than the student's estimate.  Here are two indications of the damage from this accident:
\begin{arabiclist}[(2)]
\item[(1)] Several {\it tons} of liquid helium were released, i.e.~several tons of liquid helium circulating in the LHC became gas and of course escaped; and
\item[(2)] Several of the dipole magnets, each weighing 27.5~tons, were moved by more than 10~cm.
\end{arabiclist}

\noindent
Instead of ``No beam for at least one week'', actually there was no beam for over a year.  Since this accident, low-energy proton beams did not circulate again until November 2009.


\subsection{ Gluon gluon fusion (also called gluon fusion) }\label{sec:gluon} 
As discussed in Sec.~\ref{sec:SLEP}, Higgs physics at LEP is relatively simple: an electron and a positron in the colliding beams annihilate each other into a $Z$, real or virtual, and the $Z$ is coupled to a $Z$ and an $H$ -- see Figs.~\ref{lep1} and~\ref{fig2}. The corresponding situation at the Large Hadron Collider is more complicated.

In an event at LEP, almost all the particles in the final state are seen in the detector, the major exception being the neutrinos.  In contrast, in almost any event at LHC with particle production, most of the particles in the final state are NOT seen in the detector.  In particular, many of them go down the beam pipes and are not detected.  As an example, the exclusive production process
$$p + p \rightarrow p + H + p$$
has a tiny detectable cross section.  For this reason, at the Large Hadron Collider, initially only inclusive production cross sections are of interest, at least for the Higgs particle.

To a good approximation, a proton consists of two $u$ quarks, one $d$ quark, and a number of gluons, as discussed in Sec.~\ref{sec:Introduction}.  Since the coupling of the Higgs particle to an elementary particle is proportional to its mass, there is little coupling between the Higgs particle and these constituents of the proton.  Instead, some heavy particle first needs to be produced in a proton-proton collision at the LHC, and then is used to couple to the Higgs.

From Table~\ref{tab:properties} and  Table~\ref{tab:quarks}, the heaviest known elementary particles are\\
\begin{center}
$t$,  \hspace*{3.0pt}$Z$   \hspace*{3.0pt}and   \hspace*{3.0pt} $W$ \\
\end{center}
ordered according to their masses.  In this section, we consider the case of the $t$, the top quark.

The top quark can only be produced together with an anti-top quark or an anti-bottom quark.  Since the top quark has a charge of +2/3 and is a color triplet, such pairs can be produced by

(a) a photon: $\gamma \rightarrow$ $t$ $\bar{t}$;

(b) a $Z$: $Z \rightarrow$ $t$ $\bar{t}$;

(c) a $W$: $W^{+} \rightarrow$ $t$ $\bar{b}$; or

(d) a $g$ (gluon): $g \rightarrow$ $t$ $\bar{t}$.\\
As discussed in the preceding paragraph, there is no photon, or $Z$, or $W$ as a constituent of the proton.  Since, on the other hand, there are plenty of gluons in the proton, (d) is by far the most important production process for the top quark, the heaviest known elementary particle.
     
Because of color conservation -- the gluon has color but not the Higgs particle -- the top and anti-top pair produced by a gluon cannot annihilate into a Higgs particle.  In order for this annihilation into a Higgs particle to occur, it is necessary for the top or the anti-top quark to interact with a second gluon to change its color content.  It is therefore necessary to involve two gluons, one each from the protons of the two opposing beams, and we are led to the following diagram for Higgs production: 
\begin{figure}[ht]
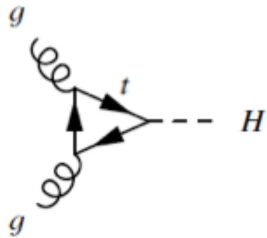

\myfig{30027f007_v2.jpg,width=1.5in}
\caption{Feynman diagram for the Higgs production by gluon gluon fusion (also called gluon fusion).}
\label{ver1}
\end{figure} \\
This production process is called ``gluon gluon fusion'' (also called ``gluon fusion'').  As expected from the large mass of the top quark, this gluon gluon fusion is by far the most important Higgs production process, and shows the central role played by the gluon\cite{Wu,Brandelik,PIEC} in the discovery of the Higgs particle in 2012.

\begin{figure}[ht]
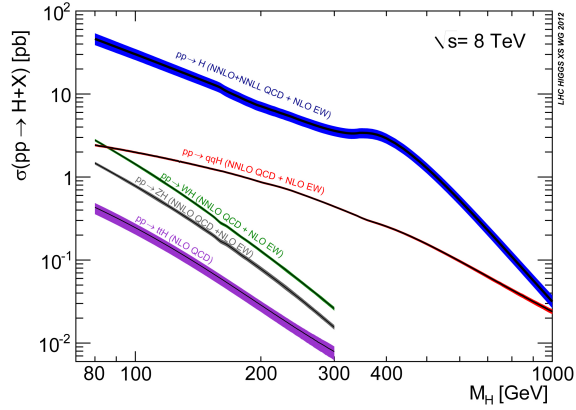

\myfig{30027f008_v2.jpg,width=3.0in}
\caption{Higgs production cross sections from gluon gluon fusion (top curve), vector boson fusion, and three associated production processes at the LHC center-of-mass energy of 8 TeV.\cite{Yellowbook}  It is seen that gluon gluon fusion is the most important production process.}
\label{ver2}
\end{figure}

It is desirable to make this statement more quantitative.  It is shown in Fig.~\ref{ver2} the various Higgs production cross sections from calculation as functions of the Higgs mass.\cite{Yellowbook}  The top curve is for gluon gluon fusion, and next one is for vector boson fusion (VBF); note that the vertical scale is logarithmic.  The other three curves are for associated production processes to be described in the next section.

\begin{figure}[ht]
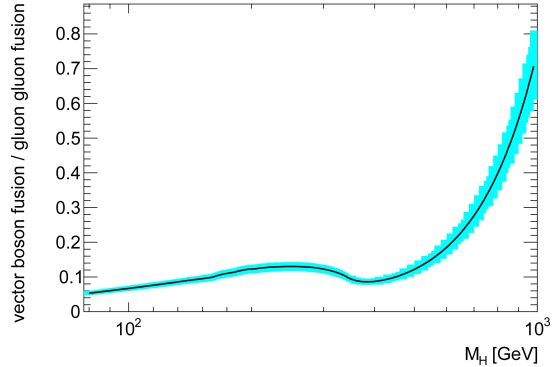

\myfig{30027f009_v2.jpg,width=3.0in}\caption{Ratio of VBF cross section divided by the gluon gluon fusion cross section.}
\label{ver3}
\end{figure}

In order to show even more clearly the importance of gluon gluon fusion, Fig.~\ref{ver3} shows the ratio of the second most important production process VBF to gluon gluon fusion.  It is seen that, for the relatively low masses of the Higgs particle, VBF cross section is less than 10\% of that of gluon gluon fusion.  It will be discussed in Sec.~\ref{sec:discovery} that the Higgs particle discovered experimentally indeed has a mass in this range. Thus, through gluon gluon fusion, the gluon contributes about ninety percent of Higgs production at the Large Hadron Collider.  A more dramatic way of saying the same thing is that, if there were no gluon, the Higgs particle could not have been discovered for many years!

\subsection{ Vector boson fusion (VBF) and associated production }\label{sec:vector}
As discussed in the preceding section, there is essentially only one way, the gluon gluon fusion, to produce the Higgs particle through its coupling to the top quark, the other contributions being much smaller.  [See, however, the comments on associated production below.]  Let us consider here the alternative possibilities through the $Z$ and the $W$ instead of the top quark.  Note, from Tables~\ref{tab:properties} and~\ref{tab:quarks}, that the mass of the $Z$ is slightly more than, while that of the $W$ slightly less than, half of the top-quark mass.
     
\begin{figure}[ht]
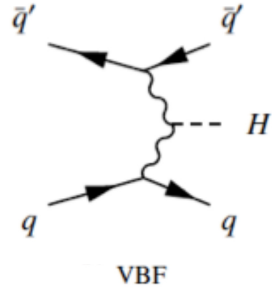

\myfig{30027f010_v2.jpg,width=1.5in}\caption{Feynman diagram for the Higgs production by vector boson fusion}
\label{ver4}
\end{figure}

Neither the $Z$ nor the $W$ has color, meaning that they do not couple to the gluon.  Thus the production of the Higgs particle through $Z/W$ is quite different from the case of Sec.~\ref{sec:gluon}.  Instead, a major process is the so-called ``Vector Boson Fusion'' (VBF) where a quark from one of the incoming protons emits a $Z$ or a $W^{+}$ while another quark from the other incoming protons provides a $Z$ or a $W^{-}$.  The $ZZ$ pair or the $W^{+} W^{-}$ pair then ``fuses'' to produce the Higgs particle.  The diagram for VBF is shown in Fig.~\ref{ver4}.

As shown in Figs.~\ref{ver2} and \ref{ver3}, the total cross section for VBF is significantly smaller than that for gluon gluon fusion, which is the major production process for the Higgs particle.  If it is desired to determine experimentally the contribution from VBF, use is made of the presence of the two quarks in the final state: each quark is seen in the detector as a $jet$.  Therefore, the Higgs particle from VBF can be seen in the detector as together with no $jet$, one $jet$, or two $jets$.

\begin{figure}[ht]
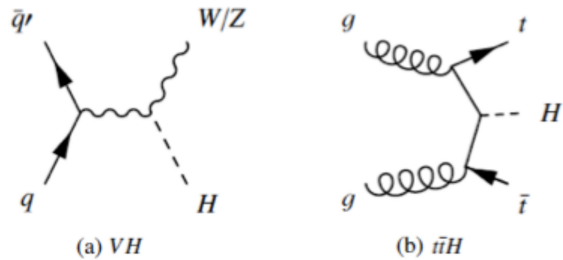

\myfig{30027f011_v2.jpg,width=3.0in}\caption{Feynman diagrams for associated Higgs production}
\label{ver5}
\end{figure}

A Higgs particle produced together with one $jet$ or two $jets$ is an example of associated production.  Three other examples of associated production are shown in Fig.~\ref{ver5}.  That of Fig.~\ref{ver5} (a) shows Higgs production in association with a $W$ or a $Z$; these processes are very similar to that of LEP in Fig.~\ref{fig2}, except that the electron-positron pair is replaced by a quark-antiquark pair.  Similarly, that of Fig.~\ref{ver5} (b) shows Higgs production in association with a top-antitop pair.  This is similar to gluon gluon fusion except that top loop is opened up; since the top quark is heavy (which is the original reason why it is desirable to use the top to produce the Higgs) the presence of one top quark and one anti-top quark in the final state reduces the production cross section for that of Fig.~\ref{ver5} (b).

All these production processes shown in Figs.~\ref{ver4} and~\ref{ver5} lead to Higgs production cross sections significantly smaller than that of gluon gluon fusion of Fig.~\ref{ver1}, the reason having already been given in the preceding section.

We now turn our attention to the decay processes for the Higgs particle.


\subsection{$H\rightarrow \gamma\gamma$}\label{sec:HtoGG}
\begin{figure}[ht]
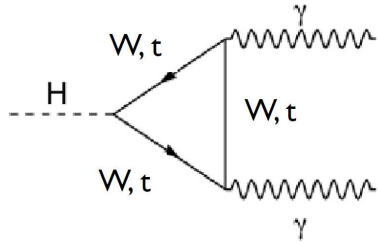

\myfig{30027f007.pdf,width=2.0in}
\caption{Feynman diagram for the decay $H\rightarrow \gamma\gamma$.}
\label{fig5}
\end{figure}

For the Higgs search, there are several important and potentially useful decay modes.  It has turned out that the most important decay mode is $H\rightarrow \gamma\gamma$.  This decay mode was fundamental in the design of the ATLAS and CMS detectors.\cite{ATLAS}
In the standard model, this decay $H\rightarrow \gamma\gamma$ proceeds predominantly through one-loop diagrams in the $W$ and the top quark, as shown in Fig.~\ref{fig5}.  For the $W$, there is an additional diagram with a $WW\gamma\gamma$ vertex.
Experimentally, this decay process has the advantage of relatively less background since the final state contains only photons.  Furthermore, because of the good energy measurement for photons with the ATLAS and the CMS detectors, this decay gives a good mass determination of $\pm 1.6~{\rm GeV}/c^2$ for the Higgs particle.

\subsection{$H\rightarrow ZZ\rightarrow\hbox{\it 4}l$}\label{sec:HtoZZ}
This is shorthand for the decay sequence
$$H\rightarrow ZZ$$
followed by$$Z\rightarrow \ell \bar{\ell}$$for both $Z$'s, where $\ell$ denotes either an electron or a muon.  Thus the final state consists of four charged leptons: $e^{+}e^{+}e^{-}e^{-}$, $e^{+}e^{-}\mu^{+}\mu^{-}$ or $\mu^{+}\mu^{+}\mu^{-}\mu^{-}$.  When the Higgs mass is less than twice the $Z$ mass, one or both of the $Z$'s may be virtual.
This channel is often referred to as the golden channel, because the final state consists of four charged leptons. Not only the momentum of each of the particles in the final state is easy to measure accurately, but also the background is under control.  However, this golden channel suffers from fairly low event rate.

\subsection{$H\rightarrow W^{+}W^{-}$} \label{sec:HtoWW}
The two processes $H\rightarrow \gamma\gamma$ and $H\rightarrow$ $ZZ\rightarrow 4l$, described in Secs.~\ref{sec:HtoGG} and \ref{sec:HtoZZ} above, are referred to as decay processes with high mass resolution.  This designation means that every one of the decay products can be measured accurately, and hence the mass of the Higgs particle can be well determined event by event.
In contrast,\break the decay mode to be considered here is:
$$H\rightarrow W^{+}W^{-}$$
followed by either:\\[3pt]
\phantom{or}\hspace*{1.5cm}both $W$'s decay leptonically,\\
or\hspace*{1.5cm}one of the $W$'s decays leptonically while the other decays hadronically.
It has not been possible to analyze the channel where both $W$'s decay hadronically, the reason being that the QCD background is too high.  When at least one of the two $W$'s decays leptonically as indicated above, there is either one or two neutrinos in the decay product, making it impossible to determine the mass of the Higgs particle event by event.  For this reason, this decay mode being considered here is referred to as a decay process with low mass resolution.
	
For some time, it was believed that this $W^{+}W^{-}$ decay mode gave the best chance of discovering the Higgs particle because of its relatively large branching ratio.  Accordingly, there was a great deal of data analysis developed for this decay mode.

\subsection{$H\rightarrow \tau^{+}\tau^{-}$ and $H\rightarrow b\bar{b}$}\label{sec:HtoTTBB}
There are two other decay channels with low mass resolution, namely, $H\rightarrow \tau^{+}\tau^{-}$  and $H\rightarrow b\bar{b}$.  Both decay modes present difficulties in their data analyses, but the difficulties are quite different.

The $\tau$ has two different types of decay modes: leptonic and hadronic. For the leptonic decay modes, there are two neutrinos in the final state; for the hadronic modes, there is one.  Therefore, for $H\rightarrow \tau^{+}\tau^{-}$, there are at least two neutrinos, making it impossible to determine the mass of the Higgs particle event by event.  On the other hand, the mass of $\tau$ is low, only 1.78~GeV/$c^{2}$ as seen from Table~\ref{tab:quarks}.  This low mass of the $\tau$ is helpful in analyzing these events.

For Higgs mass below about 135~GeV/$c^{2}$, the branching fraction for the decay $H\rightarrow b \bar{b}$ is large.  Unfortunately, the signal for this decay is overwhelmed by the QCD production of bottom quarks.  Therefore, the search in this channel is limited to associated production, i.e.,~where the Higgs particle is produced together with a $W$ or a $Z$, as discussed in Sec.~\ref{sec:vector}.  Specifically, the channels that have been studied consist~of:
$$ZH\quad\mbox{with}\quad H\rightarrow b\bar{b}\quad\mbox{and}\quad
Z\rightarrow \ell^{+}\ell^{-}\ \mbox{or}\ \nu \bar{\nu}\,,$$
and
$$W^{-}H\quad\mbox{with}\quad H\rightarrow b\bar{b}\quad\mbox{and}\quad W^{-}\rightarrow  \ell^{-} \bar{\nu}$$
(together with $W^{+}H$).

\baselineskip=12.9pt

\section{Discovery!}\label{sec:discovery}
\subsection{Operation of the Large Hadron Collider}\label{sec:oplhc}
As described in Sec.~\ref{sec:setback}, after the major setback in September 2008, the LHC began to operate again in November 2009.  For the year 2010, this LHC operated at the center-of-mass energy of 7~TeV, half of the design energy.

The original plan was to continue to run at this energy of 7~TeV until the end of 2011, at which time a long technical stop of a couple of years began in order to prepare the Collider for running at its full design energy of 7~TeV per beam.  On January~31, 2011, CERN announced a revision of this plan.  Instead, the LHC would run through the end of 2012 with a short technical stop at the end of 2011.  The long technical stop to reach the full design energy will take place one year later in 2013 to 2014.

This revision of the original plan has turned out to be crucial for the discovery of the Higgs particle, to be described in this section.  If the original plan had been followed, this Higgs particle certainly would not have been discovered yet.

Because of the excellent performance of the Large Hadron Collider in the year 2011, the CERN management made another excellent decision to increase the center-of-mass energy from 7~TeV for 2011 to 8~TeV for 2012.  This increase in energy was also most helpful to the discovery of the Higgs particle in 2012.

In this section, the exciting events, mostly in the years 2011 and 2012, are described in chronological order, culminating in the discovery of the Higgs particle, announced on July 4, 2012.

\subsection{The year $2011$}\label{sec6.2}
No experimental indication of the Higgs particle was seen in the first half of 2011.  During the summer of 2011, at the EPS-HEP 2011 in July\cite{EPS} and at the Lepton--Photon 2011 in August\cite{Lepton}, both the ATLAS Collaboration and the CMS\break Collaboration saw first indications of a Higgs particle: $2.1\sigma$ at 145~GeV/$c^{2}$ from ATLAS and $2.3\sigma$ at 120~GeV/$c^{2}$ from CMS.  These indications both came from $H\rightarrow W^{+}W^{-}$ with the $W$'s decaying leptonically.  This is the decay mode described in Sec.~\ref{sec:HtoWW}, a decay process with low mass resolution.  Thus the masses of 145~GeV/$c^{2}$ and 120~GeV/$c^{2}$ are not considered to be in disagreement, and such masses are referred to as ``low mass''.

Encouraged by this indication, attention then turned to the two Higgs decay processes with high mass resolution: $H\rightarrow \gamma\gamma$ and $H\rightarrow ZZ\rightarrow 4l$ described in Secs.~\ref{sec:HtoGG} and \ref{sec:HtoZZ} above.

\baselineskip=13pt

Less than half a year later, at the CERN Council meeting on December~13, 2011, both the ATLAS Collaboration and the CMS Collaboration showed signals of more than $2\sigma$  in both of the above decay processes at closeby Higgs masses. Figures~\ref{fig6} and \ref{fig7} show the published versions\cite{ATLASPUBFig8,CMSPUBFig8,ATLASPUBFig9, CMSPUBFig9} of the then preliminary plots shown at this council meeting.  In these figures, as well as in later ones, $p_{0}$ or $p$-value means the probability that the background fluctuates to the observed data (or higher).

Note that the four masses of lowest $p_{0}$ in Figs.~\ref{fig6} and \ref{fig7} are all very close, well within 10~GeV/$c^{2}$ of each other.

It was at this point when a major excitement started building up.

\begin{figure}[ht]
\myfig{30027f008.pdf,width=5.0in}
\caption{Background fluctuation probability $p_0$ in the $H \rightarrow \gamma\gamma$ channel, using the full 2011 datasets collected by ATLAS (left) and CMS (right). The Gaussian significances corresponding to the peaks in $p_0$ near $m_H=125$~GeV are also shown. (ATLAS: Phys. Rev. Lett. 108, 111803 (2012)~\cite{ATLASPUBFig8}; CMS: Phys. Lett. B 710 (2012) 403-425\cite{CMSPUBFig8}) The corresponding preliminary plots shown on December 13, 2011 are given in these links: \protect\url{https://atlas.web.cern.ch/Atlas/GROUPS/PHYSICS/CONFNOTES/ATLAS-CONF-2011-161/fig_07.png} (ATLAS) and \protect\url{http://cds.cern.ch/record/1406346/files/Figure_004.pdf} (CMS).}
\label{fig6}
\vspace*{4pt}

\myfig{30027f009.pdf,width=5.0in}
\caption{Background fluctuation probability $p_0$ in the $H \rightarrow ZZ\rightarrow 4l$ channel, using the full 2011 datasets collected by ATLAS (left) and CMS (right). The Gaussian significances corresponding to the peaks in $p_0$ near $m_H=125$~GeV are also shown. (ATLAS: Phys. Lett. B710 (2012) 383-402~\cite{ATLASPUBFig9}; CMS: Phys. Rev. Lett. 108, 111804 (2012)\cite{CMSPUBFig9}) The corresponding preliminary plots made public on December 13, 2011 are given in these links: \protect\url{https://atlas.web.cern.ch/Atlas/GROUPS/PHYSICS/CONFNOTES/ATLAS-CONF-2011-162/fig_07b.png} (ATLAS) and \protect\url{http://cds.cern.ch/record/1406342/files/Figure_002-a.pdf} (CMS).}
\label{fig7}
\end{figure}

\begin{figure}[ht]
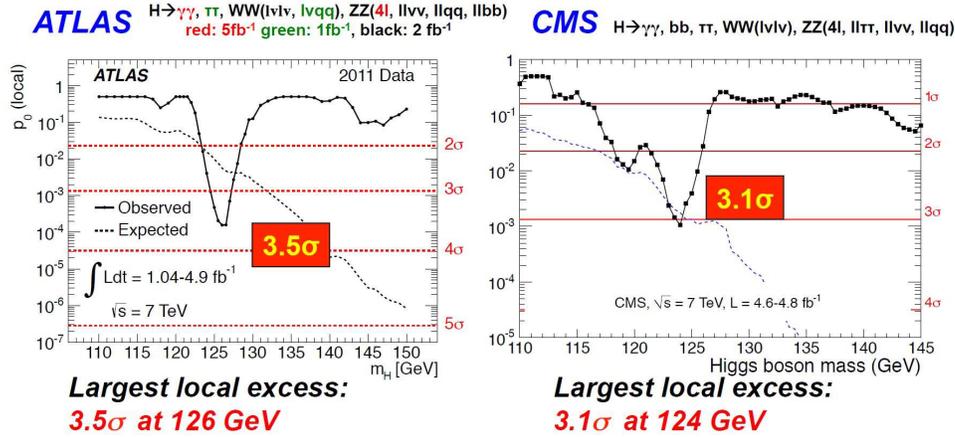

\myfig{30027f010.pdf,width=5in}
\caption{Background fluctuation probability $p_0$ for Higgs search by ATLAS (left) and CMS (right), using a combination of all available search channels. The ATLAS results used between 1~f\hspace*{0.04em}b$^{-1}$ and 5~f\hspace*{0.04em}b$^{-1}$ of the 2011 dataset depending on the channel, while the CMS results used the full 2011 dataset in each channel. The Gaussian significances corresponding to the peaks in $p_0$ are also shown. (ATLAS: Phys. Lett. B 710 (2012) 49-66~\cite{ATLASPUBFig10}; CMS: Phys. Lett. B 710 (2012) 26-48\cite{CMSPUBFig10}) The corresponding preliminary plots shown on December 13, 2011 are given in these links: \protect\url{https://atlas.web.cern.ch/Atlas/GROUPS/PHYSICS/CONFNOTES/ATLAS-CONF-2011-163/fig_06b.png} (ATLAS) and \protect\url{http://cms-higgs-results.web.cern.ch/cms-higgs-results/Comb/HIG-11-032/pvala_comb_comp_zoom.pdf} (CMS).}
\label{fig8}
\end{figure}

When all the decay channels are combined, the results for the ATLAS Collaboration and the CMS Collaboration were presented in the CERN Council Meeting on December~13, 2011. Again, Fig.~\ref{fig8} shows the published versions\cite{ATLASPUBFig10,CMSPUBFig10} of the preliminary plots shown at this council meeting.

\subsection{June $2012$}
With the results presented at the above-mentioned CERN Council Meeting, there was then a small mass range for the ATLAS Collaboration and the CMS Collaboration to hunt for the Higgs particle. There was frantic activity for the next six months with new data coming in.  As typical of data analysis in high-energy physics, the detailed method of analysis was decided first on the basis of Monte Carlo simulation without looking at the actual data: this is to avoid mistaking accidental random fluctuation as real signal.  It should perhaps be mentioned that the data analysis was quite complicated, involving digging out the pertinent events from the enormous amount of data being recorded.

After the detailed method of analysis was decided on, the actual experimental data were made available to carrying out the analysis --- this process is referred to as unblinding.  Because of the decision to continue operating the Large Hadron Collider in 2012 instead of turning it off at the end of 2011 (see Sec.~\ref{sec:oplhc}), there was a rapid accumulation of additional data since the 2011 CERN Council Meeting on \hbox{December~13,} and the unblinding was scheduled for June 2012.  This date of unblinding was determined by the date of the International Conference on High-Energy Physics beginning on July~5, in Australia.

The CMS physicists gathered on June 15 in Room 222 of the CERN filtration plant to hear the first report after the unblinding.\cite{scientific}  The first speaker talked about the decay $H\rightarrow W^{+}W^{-}$, the channel where the first indication for the Higgs particle was seen in the summer of 2011.  In this channel, a small excess was seen in the 2012 data, but this small signal did not generate any excitement.  This talk was followed by two presentations respectively on $H\rightarrow \gamma\gamma$ and $H\rightarrow ZZ \rightarrow 4l$, the two decay processes with high mass resolution; see Sec.~\ref{sec:HtoWW}.  For both of these two decay channels, the signals from the LHC data of 2012 were found to occur again in the vicinity of 125 GeV/c$^{2}$ seen six months earlier from the 2011 data.  See Figs.~\ref{fig6} and~\ref{fig7}. The crowd cheered enthusiastically at the end of these two presentations.

Similar revelations occurred in the ATLAS Collaboration: spontaneous celebrations\- broke out in several groups when the data were first unblinded.  These celebrations were followed by more than a week of long workdays and sleepless nights in order to combine the signals from the 2011 and 2012 data (which were taken at 7~TeV and 8~TeV, respectively) and also to combine the different channels. In the afternoon of June 25, 2012 during a meeting, a graduate student presented the first successful combination of the data, leading to a result of 5.08$\sigma$ giving a Higgs mass of 126.5 GeV/c$^{2}$ (See Sec.~\ref{sec:july4} below). This presentation caused cheers ringing down a corridor. Another graduate student very soon confirmed this result independently.

Why is this $5\sigma$ signal so very important?  In particle physics, $5\sigma$ is considered to be the golden standard to claim a discovery; it means that the probability for the background to fluctuate to the observed data or higher is less than one in three million.

\subsection{July~$4,2012$ -- a great day for physics}\label{sec:july4}
The original plan was to make this discovery of the Higgs particle public on July~5 in Australia, the first day of the 2012 International Conference on High-Energy Physics.  Since the experiments are sited at CERN, the decision was made instead to announce this result from the ATLAS Collaboration and the CMS Collaboration\- one day earlier at CERN.  Thus a special symposium was arranged in the CERN auditorium on July~4, which happens to be the US Independence\break Day.

With the auditorium locked, physicists and physics students slept just outside of the auditorium the night before. Fran\c{c}ois Englert, Peter Higgs, Gerald Guralnik and Carl Hagen\cite{Englert} walked into the auditorium in the morning of July~4 to a standing ovation.  Robert Brout, who had contributed a great deal to this phenomenon from the theoretical side, had unfortunately passed away.
\begin{figure}[ht]
\myfig{30027f011.pdf,width=5.0in}
\caption{Background fluctuation probability $p_0$ in the $H \rightarrow \gamma\gamma$ channel, using the full 2011 datasets and partial 2012 datasets collected by ATLAS (left) and CMS (right). The Gaussian significances corresponding to the peaks in $p_0$ are also shown.\cite{atlasdisc,cmsdisc} The corresponding preliminary plots shown on July 4, 2012 are given in these links: \protect\url{https://atlas.web.cern.ch/Atlas/GROUPS/PHYSICS/CONFNOTES/ATLAS-CONF-2012-091/fig_10.png} (ATLAS) and \protect\url{https://twiki.cern.ch/twiki/pub/CMSPublic/Hig12015TWiki/MVApvaluescomb.pdf} (CMS).}
\label{fig9}
\myfig{30027f012.pdf,width=5.0in}
\caption{Background fluctuation probability $p_0$ in the $H \rightarrow ZZ \rightarrow 4l$ channel, using the full 2011 datasets and partial 2012 datasets collected by ATLAS (left) and CMS (right). The Gaussian significances corresponding to the peaks in $p_0$ are also shown.\cite{atlasdisc, cmsdisc} The corresponding preliminary plots shown on July 4, 2012 are given in these links: \protect\url{https://atlas.web.cern.ch/Atlas/GROUPS/PHYSICS/CONFNOTES/ATLAS-CONF-2012-092/fig_15a.png} (ATLAS) and \protect\url{https://twiki.cern.ch/twiki/pub/CMSPublic/Hig12016TWiki/Pvals_PLP_lowMass_2D_no2l2tau_7p8sep.png} (CMS).}
\label{fig10}
\end{figure}

\begin{figure}[tb]
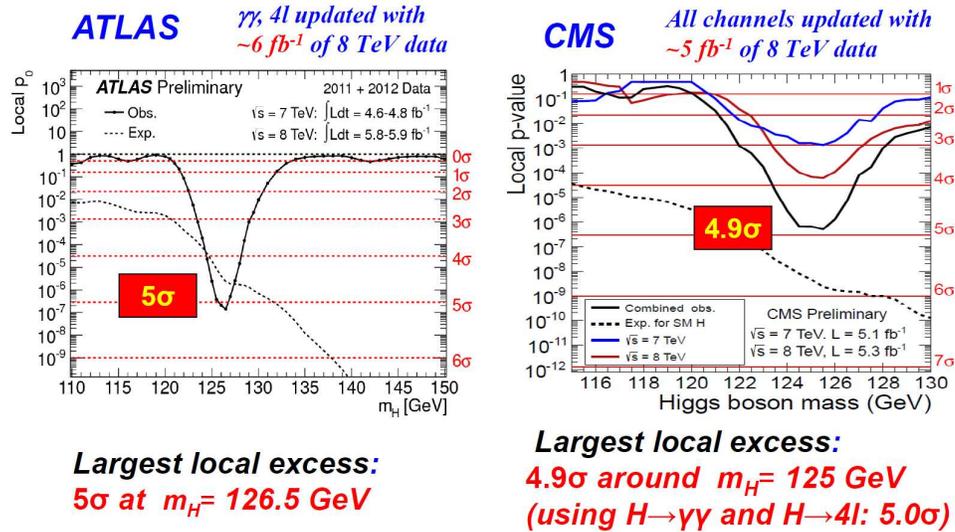

\myfig{30027f013.pdf,width=5.0in}
\caption{Background fluctuation probability $p_0$ in the Higgs search using the full 2011 datasets and partial 2012 datasets collected, as shown in the CERN special symposium by the ATLAS Collaboration and CMS Collaboration on July~4, 2012. The Gaussian significance of $5\sigma$ corresponding to the peak in $p_0$ is indicated. These preliminary plots shown on July 4, 2012 are given in these links: \protect\url{https://atlas.web.cern.ch/Atlas/GROUPS/PHYSICS/CONFNOTES/ATLAS-CONF-2012-093/fig_03b.png} (ATLAS) and \protect\url{http://cms-higgs-results.web.cern.ch/cms-higgs-results/Comb/HIG-12-020//sqr_pvala_all_energy.pdf} (CMS).}
\label{fig11}
\end{figure}

The published version of the results presented at this special symposium are shown in  Figs.~\ref{fig9}--\ref{fig10} for $H\rightarrow \gamma\gamma$ and $H\rightarrow ZZ \rightarrow 4l$. Again when all the decay channels are combined, the results for the ATLAS Collaboration and the CMS Collaboration were presented in the CERN special symposium on July 4, 2012. Figure~\ref{fig11} shows the preliminary plots of ATLAS and CMS presented at this CERN special symposium. The same figures were shown in the ICHEP conference in Melbourne, Australia, on July 4-11, 2012. Both ATLAS and CMS showed their final significance, obtained by combining all their decay channels, with ATLAS obtaining $5\sigma$ and CMS $4.9\sigma$. 

Right after the ATLAS and CMS presentations, Rolf Heuer, the Director General of CERN declared ``I think we have it. [...] We have now found the missing cornerstone of particle physics. We have a discovery. We have observed a new particle that is consistent with a Higgs boson.''

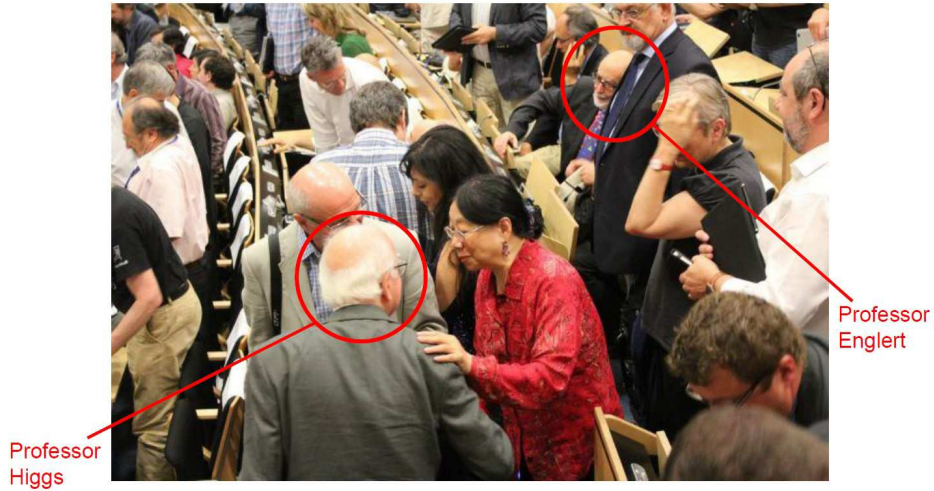
A scene in the auditorium is shown in Fig.~\ref{fig12}.  At the end of this special symposium, The author went to shake hands with Peter Higgs.  I told him ``I have been looking for you for over 20~years''.  He replied ``now, you have found me'' --- Fig.~\ref{fig13}.

After this special symposium, the ATLAS Higgs working group had a celebratory drink --- Fig.~\ref{fig14}.  Everyone was watching the corks of the champagne bottles flying up to the ceiling.

\begin{figure}[ht]
\myfig{30027f014.pdf,width=4.8in}
\caption{Scenes of jubilation in the CERN auditorium on July~4, 2012. Several former Director Generals of CERN are seen in this photograph. Standing, from left to right: Christopher Llewellyn Smith (1994-1998), Lyn Evans (leader of the LHC Project), Herwig Schopper (1981-1988), Luciano Maiani (1999-2003) and Robert Aymar (2004-2008). The present CERN Director General Rolf Heuer is at the podium, facing the audience, not seen in this picture. (Photo from New York Times on July 4, 2012 by Denis Balibouse)}
\label{fig12}
\vspace*{4pt}
\myfig{30027f015.pdf,width=4.9in}
\caption{The author met Prof. Higgs.}
\label{fig13}
\end{figure}

\begin{figure}[ht]
\myfig{30027f016.pdf,width=5.0in}
\caption{Celebration by the ATLAS Higgs working group following the discovery announcements on July~4, 2012. (Photo by Dr. Natalia Panikashvili)}
\label{fig14}
\end{figure}

\subsection{First publications}\label{sec:sec6.5}
On July~31, 2012, both the ATLAS Collaboration and the CMS Collaboration submitted their first publications on the discovery of the Higgs particle to the Physics Letters B.  In the ATLAS publication, the analysis for the decay channel $H\rightarrow W^{+}W^{-}$ was updated (with respect to that included in the result shown in the July 4 CERN seminar), leading to a $5.9\sigma$ at the mass of 126.5~GeV/$c^{2}$. These two publications are:

\vspace*{5pt}
ATLAS Collab., {\it Phys. Lett. B} {\bf 716}, 1 (2012)\,,

CMS Collab., {\it Phys. Lett. B} {\bf 716}, 30 (2012)\,.\\[5pt]
Both the published result of the ATLAS Collaboration of $5.9\sigma$ and the published result of CMS Collaboration of $5\sigma$ are given in Fig.~\ref{fig15}\cite{atlasdisc, cmsdisc}. The cover of this issue dated September~17, 2012 is shown as Fig.~\ref{fig16}; there are two figures on this cover, the upper one from CMS and the lower one from ATLAS.
\begin{figure}[ht]
\myfig{30027f017.pdf,width=5.0in}
\caption{Background fluctuation probability $p_0$ in the Higgs search using the full 2011 datasets and partial 2012 datasets collected by ATLAS (left) and CMS (right), as shown in the discovery publication submitted by the ATLAS and the CMS Collaborations. The results were obtained by combining all available standard model decay channels. The Gaussian significances corresponding to the peak in $p_0$ is $5.9\sigma$ for ATLAS\cite{atlasdisc} and $5\sigma$ for CMS.\cite{cmsdisc} }
\label{fig15}

\myfig{30027f018.pdf,width=7cm,height=9cm}
\caption{Cover of the September~17, 2012 issue of {\it Phys. Lett.~B}, showing some of the Higgs results from the ATLAS and CMS experiments.}
\label{fig16}
\end{figure}
A first test of the compatibility of the observed particle with the Higgs particle is provided by the signal
strength $\mu = \sigma/\sigma_{\rm SM}$ from the different decay channels. Here $\sigma$ is the observed cross-section and $\sigma_{\rm SM}$ is the corresponding prediction on the basis of the standard model. The ATLAS and the CMS results, obtained with the data used for the discovery, are shown in Fig.~\ref{fig19}. It is seen that, from both Collaborations, the important decay channels are $H\rightarrow \gamma\gamma$, $H\rightarrow ZZ \rightarrow 4l$ and $H\rightarrow WW$, discussed in Secs.~\ref{sec:HtoGG}, \ref{sec:HtoZZ} and \ref{sec:HtoWW}, respectively. It should be noted from Figs.~\ref{ver2} and~\ref{ver3} that, at this Higgs mass, gluon gluon fusion is responsible for about 90\% of its production, as discussed in Sec.~\ref{sec:gluon}.

\begin{figure}[t]
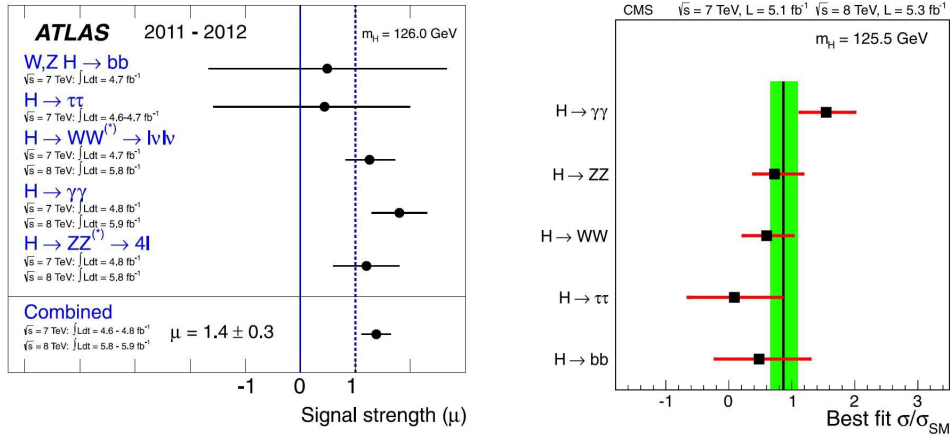

\myfig{30027f019.pdf,width=5in}
\caption{Measurements of the signal strength \protect$\mu = \sigma/\sigma_{\rm SM}$ for the individual channels and their combinations in July 2012 in ATLAS\cite{atlasdisc} (left) and CMS\cite{cmsdisc} (right).}
\label{fig19}
\vspace*{2pt}
\end{figure}

The results presented in Fig.~\ref{fig19} have been updated with the entire data set of 25~f\hspace*{0.06em}b$^{-1}$ taken by the ATLAS and the CMS Collaborations before the shut down of the Large Hadron Collider in 2013 (the updated results are not shown here).

\enlargethispage*{13pt}

\subsection{Spin-parity}\label{sec6.6}
When this new particle was discovered in the summer of 2012, it was cautiously referred to as a ``Higgs-like boson''.  As discussed in Sec.~\ref{sec:higgsparticle}, the most fundamental property of the Higgs particle is that it must have spin zero. Therefore, a first priority was to determine the spin of this new particle.

For $H\rightarrow ZZ\rightarrow 4l$, spin-parity analyses have been performed for masses near 125~GeV/$c^{2}$. Moreover, spin analyses have been performed in the  $H\rightarrow WW\rightarrow lvlv$ and the  $H\rightarrow \gamma\gamma$ channels. Independently, both the ATLAS Collaboration and the CMS Collaboration have found that the $0^{+}$ state is favored over the $0^{-}$, $1^{+}$, $1^{-}$ and $2^{+}$ states at 95\% confidence level or better.\cite{ACPUBSpin} By now, it is generally accepted that this newly found particle is indeed the Higgs particle.

\section{The Nobel Prize in Physics 2013}\label{sec:nobel}
\begin{figure}[t]
\myfig{30027f021.pdf,width=3in}
\caption{}
\label{fig21}
\vspace*{2pt}
\end{figure}

On October 8, 2013, the Royal Swedish Academy of Sciences announced to award the Nobel Prize in Physics for 2013 to Fran\c{c}ois Englert (Universit\'e Libre de Bruxelles, Brussels, Belgium) and Peter W. Higgs ( University of Edinburgh, UK) -Fig.~\ref{fig21}.

The citation is \\

\noindent \setlength{\leftskip}{2em}\setlength{\rightskip}{2em}``for the theoretical discovery of a mechanism that contributes to our understanding of the origin of mass of subatomic particles, and which recently was confirmed through the discovery of the predicted fundamental particle, by the ATLAS and CMS experiments at CERN's Large Hadron Collider'' \\ 
\\ \\

\setlength{\leftskip}{0pt}\setlength{\rightskip}{0pt}
\noindent and the press release is \\
  
\noindent \setlength{\leftskip}{2em}\setlength{\rightskip}{2em}``Fran\c{c}ois Englert and Peter W. Higgs are jointly awarded the Nobel Prize in Physics 2013 for the theory of how particles acquire mass. In 1964, they proposed the theory independently of each other (Englert together with his now deceased colleague Robert Brout). In 2012, their ideas were confirmed by the discovery of a so called Higgs particle at the CERN laboratory outside Geneva in Switzerland.'' \\

\setlength{\leftskip}{2pt}\setlength{\rightskip}{0pt}
This prize is well deserved.

\section{Looking Forward}\label{sec7}
Before this discovery of the Higgs particle, all the elementary particles have either spin~1 or spin~$\frac{1}{2}$; see Tables~\ref{tab:properties} and \ref{tab:quarks}.  Since the Higgs particle has spin~0, this discovery is not only of a new elementary particle, but also a new type of elementary particle.

For this reason, many particle physicists consider this discovery to be the most important step forward in particle physics for the last half a century: it opens up a new regime and therefore a new era of physics.  What can we say about this new era of physics?

\baselineskip=13.2pt

It is the purpose of this section to consider some of the possibilities.

\enlargethispage{-10pt}

Before embarking on these speculations, it may be instructive to recall an event of more than 60~years ago.  In the '40s, the pion was discovered.  It was the particle that Yakawa had proposed as the mediator of nuclear interactions.  But we now know much better:  instead of being just the particle for nuclear interactions, this pion is in fact the first one of a large class of particles called mesons.  Soon after its discovery, the
K~(kaon) was found in cosmic rays.

If this analog has any relevance, then we may consider the attractive possibility that the Higgs particle is not the only one of the spin-0 elementary particles.

Do we have any handle on this possibility?

As seen in Fig.~\ref{fig19} from the first publications of the ATLAS Collaboration and the CMS Collaboration, the channel $H\rightarrow \gamma\gamma$ gives a higher event rate compared with the predictions of the standard model.\cite{Glashow}

This decay is also a most interesting process from the theoretical point of view.\cite{Yellowbook} Since the photon has no mass, it does not couple directly to the Higgs particle in the standard model: the decay proceeds predominantly through a top or a $W$ loop, as discussed in Sec.~\ref{sec:HtoGG} and Fig.~\ref{fig5}. These one-loop contributions to this decay must be finite, and therefore can be calculated in a completely straightforward way without introducing complications such as regularization or ghosts.

\begin{figure}[t]
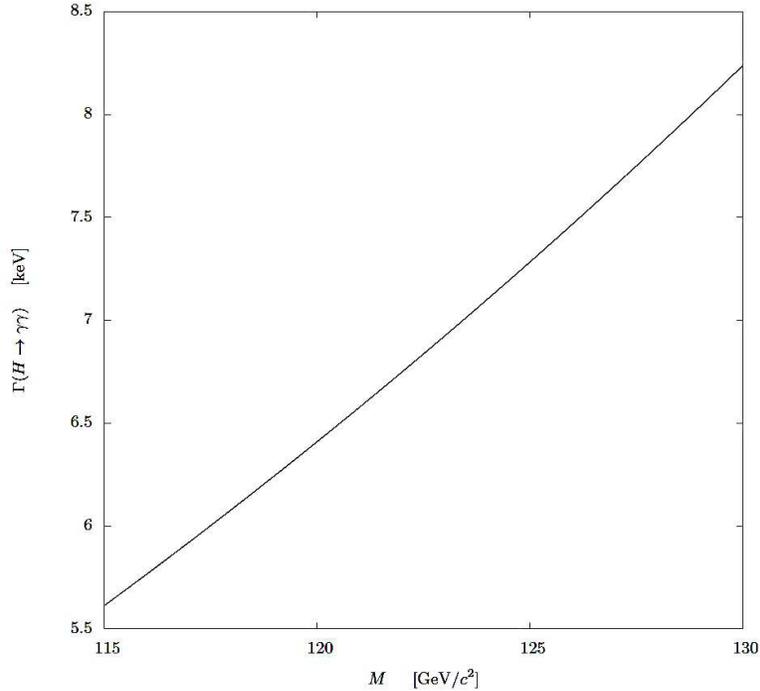

\myfig{30027f020.pdf,width=4in}
\caption{The decay width $\Gamma (H\rightarrow \gamma\gamma)$ as a function of the Higgs boson mass $M$.\cite{Gastman} }
\label{fig20}
\end{figure}

\baselineskip=13pt

Calculations have been carried out as discussed in Ref.~\refcite{Yellowbook} and references herein. In addition another calculation shows a more extensive cancellation of the contribution between the top and $W$.\cite{Gastman,Rizzo} The result for this decay $H\rightarrow \gamma\gamma$ is shown in Fig.~\ref{fig20}, implying that the signal strengths shown in Fig.~\ref{fig19} for $H\rightarrow \gamma\gamma$ have been underestimated; instead the experimentally observed decay rate is about a factor of~3 above the expectations from the standard model\cite{Gastman}.

If this excess remains with additional data from the Large Hadron Collider at 14~TeV, then the most likely explanation is that there are new heavy charged particles that contribute to the decay $H\rightarrow \gamma\gamma$ in addition to the top and the $W$.  This would be most exciting: the experimental observation of the Higgs particle not only completes the list of particles for the standard model\cite{Glashow}, but also gives a first indication of the physics beyond the standard model!

\section*{Acknowledgments}
The discovery of the Higgs particle, announced on July~4, 2012, was the result of two decades of work by the ingenious LHC machine physicists and by many thousands of ATLAS and CMS experimental physicists who built and now operate the detectors, designed and now manage a computer system that distributes data around the world, created novel hardware and computer software to identify the most interesting collisions, and wrote the algorithms that dig out the most pertinent events from the great morass of data being recorded. They all worked feverishly, contributing greatly to the Higgs discovery. 

I thank the hospitality and support of CERN where I spent a large fraction of my research time since 1985. 
I thank many of my ATLAS colleagues for most exciting interaction and stimulating discussion. I thank some of the CMS colleagues for their most inspiring spirit. I am grateful to many recent members of my Wisconsin group who have worked very hard and over long periods to contribute to the ATLAS discovery of the Higgs particle; they include Swagato Banerjee, Elizabeth Castaneda, Luis Flores Castillo, Yaquan Fang, Andrew Hard, Yang Heng, Richard Jared, Haoshuang Ji, John Joseph, Xianyang Ju, Lashkar Kashif, Haifeng Li, Lianliang Ma, Yao Ming, German Carrillo Montoya, Isabel Pedraza, William Quayle, Fuquan Wang, Haichen Wang, Werner Wiedenmann, Hongtao Yang, Fangzhou Zhang and Haimo Zobernig.  Together with our ATLAS collaborators, they contributed to the ATLAS Discovery through their work on $H\rightarrow \gamma\gamma$, $H\rightarrow$ $ZZ\rightarrow 4l$, $H\rightarrow WW$, $H\rightarrow \tau\tau$, $H\rightarrow bb$, Higgs decay channels combination and High Level triggers.  I am also indebted to many of the former members of my group for their important contributions, but there are too numerous to be listed here. I thank their support at difficult times.

I am most grateful to the support, throughout many years, of the United States Department of Energy (Grant No. DE-FG02-95ER40896) and the University of Wisconsin through the Wisconsin Alumni Research Foundation and the Vilas Foundation.


\begin{thebibliography}{34}
\bibitem{Einstein} \afm{A.} \asm{Einstein}, \at{On a heuristic point of view concerning the production and transformation of light}, \jat{Ann. Phys.} \vol{{\bf 17}}, \pgs{132} (\y{1905}).

\bibitem{Yang} \afm{C. N.} \asm{Yang} and \afm{R. L.} \asm{Mills}, \jat{Phys. Rev.} \vol{{\bf 96}}, \pgs{191} (\y{1954}).

\bibitem{Wu} \ccm{S. L. Wu and G. Zobernig, {\it Z. Phys. C$:$ Part. Fields} {\bf 2}, 107 (1979); TASSO Note 84, June~26, 1979}.

\bibitem{Brandelik}\ccm{R. Brandelik {\it et~al.} (TASSO Collab.), {\it Phys. Lett. B} {\bf 86}, 243 (1979); D. P. Barber {\it et~al.} (MARK J Collab.), {\it Phys. Rev. Lett.} {\bf 43}, 830 (1979); Ch. Berger {\it et~al.} (PLUTO Collab.), {\it Phys. Lett. B} {\bf 86}, 418 (1979); W. Bartel {\it et~al.} (JADE Collab.), {\it ibid.} {\bf 91}, 142 (1980)}.

\bibitem{PIEC}\ccm{P. S\"{o}ding, B. Wiik, G. Wolf  and S. L. Wu (presented by S. L. Wu) for the 1995 EPS High Energy and Particle Physics Prize,  Proceedings of the International Europhysics Conference on High Energy Physics, Brussels, Belgium 27 Jul - 2 Aug 1995, pp. 3-10}.

\bibitem{Arnisson} \ccm{G. Arnisson {\it et~al.} (UA1 Collab.), {\it Phys. Lett. B} {\bf 122}, 103 (1983); {\it ibid.} {\bf 126}, 398 (1983)}.

\bibitem{Banner} \ccm{M. Banner {\it et~al.} (UA2 Collab.), {\it Phys. Lett. B} {\bf 122}, 476 (1983); {\it ibid.} {\bf 129}, 130 (1983)}.

\bibitem{phys} \org{Particle Data Group}, \jat{Phys. Rev. D} \vol{{\bf 86}}, \pgs{1} (\y{2012}).

\bibitem{GellMann} \afm{M.} \asm{Gell-Mann}, \jat{Phys. Lett. B} \vol{{\bf 8}}, \pgs{241} (\y{1964}).

\bibitem{Zweig} \ccm{G. Zweig, CERN Report 8182/TH-401 (1964); CERN Report 8419/TH-412 (1964)}.

\bibitem{Glashow} \ccm{S. Glashow, {\it Nucl. Phys.} {\bf 22}, 579 (1961); S. Weinberg, {\it Phys. Rev. Lett.} {\bf 19}, 1264 (1967); A. Salam, {\it Proc. 8th. Nobel Symp.} (Stockholm, 1968), ed. N. Svartholm (Almqvist, 1968), p.~367}.

\bibitem{Englert} \ccm{F. Englert and R. Brout, {\it Phys. Rev. Lett.} {\bf 13}, 321 (1964); P. W. Higgs, {\it Phys. Lett.} {\bf 12}, 132 (1964); P.~W. Higgs, {\it Phys. Rev. Lett.} {\bf 13}, 508 (1964); G.~S. Guralnik, C.~R. Hagen and T.~W.~B. Kibble, {\it ibid.} {\bf 13}, 585 (1964); P.~W. Higgs, {\it Phys. Rev.} {\bf 145}, 1156 (1966); T.~W.~B. Kibble, {\it ibid.} {\bf 155}, 1554 (1967)}.

\bibitem{atlasdisc} \org{The ATLAS Collab.}, \jat{Phys. Lett. B} \vol{{\bf 716}}, \pgs{1} (\y{2012}).

\bibitem{cmsdisc} \org{The CMS Collab.}, \jat{Phys. Lett. B} \vol{{\bf 716}}, \pgs{30} (\y{2012}).

\bibitem{Fermilab} \ccm{F. Abe {\it et~al.} (CDF Collab.), {\it Phys. Rev. Lett.} {\bf 74}, 2626 (1995); S. Abachi {\it et~al.} (D0 Collab.), {\it ibid.} {\bf 74}, 2632 (1995)}.

\bibitem{Crystal} \afm{H. J.} \asm{Trost} (\org{for the Crystal Ball Collab.}), \pt{{\it Proc. 22nd Int. Conf. on High-Energy Physics}}, Vol.~\vol{1}, \cp{July~19--25, 1984, Leipzig, Germany}, eds. \efm{A.} \esm{Meyer} and \efm{E.} \esm{Wieczorek}, \pp{Zeuthen, East Germany} (\pn{Akad. Wiss.}, \y{1984}), pp.~\pgs{201--203}.

\bibitem{CUSB} CUSB Collab., \pt{{\it Proc. of the XXIV Int. Conf. on High-Energy Physics}}, \cp{Munich, 1989}), p.~\pgs{891}.

\bibitem{SINDRUM} \org{SINDRUM Collab.}, \jat{Phys. Lett. B} \vol{{\bf 222}}, \pgs{533} (\y{1989}).

\bibitem{CLEO} \org{CLEO Collab.}, \jat{Phys. Rev. D} \vol{{\bf 40}}, \pgs{712} (\y{1990}).

\bibitem{Mermikides} \ccm{M. Mermikides, S. L. Wu and H. Zobernig, ALEPH note, December~21, 1983}.

\bibitem{LEP} \ccm{ALEPH Collab., {\it Phys. Lett. B} {\bf 236}, 233 (1990); {\it ibid.} {\bf 237}, 291 (1990); {\it ibid.} {\bf 241}, 141 (1990); {\it ibid.} {\bf 245}, 289 (1990); {\it ibid.} {\bf 246}, 306 (1990); {\it ibid.} {\bf 313}, 299 (1993); {\it ibid.} {\bf 384}, 427 (1996); DELPHI Collab., {\it Nucl. Phys. B} {\bf 342}, 1 (1990); DELPHI Collab., {\it Phys. Lett. B} {\bf 245}, 276 (1990); DELPHI Collab., {\it Z. Phys. C} {\bf 51}, 25 (1991); DELPHI Collab., {\it Nucl. Phys. B} {\bf 373}, 3 (1992); {\it ibid.} {\bf 421}, 3 (1994); L3 Collab., {\it Phys. Lett. B} {\bf 248}, 203 (1990); {\it ibid.} {\bf 252}, 518 (1990); {\it ibid.} {\bf 257}, 450 (1991); {\it ibid.} {\bf 283}, 454 (1992); {\it ibid.} {\bf 303}, 391 (1993); {\it ibid.} {\bf 385}, 454 (1996); OPAL Collab., {\it Phys. Lett. B} {\bf 236}, 224 (1990); {\it ibid.} {\bf 251}, 211 (1990); OPAL Collab., {\it Z. Phys. C} {\bf 49}, 1 (1991); OPAL Collab., {\it Phys. Lett. B} {\bf 268}, 122 (1991); {\it ibid.} {\bf 253}, 511 (1991); {\it ibid.} {\bf 327}, 397 (1994)}.

\bibitem{SauLanWu} \pti{Search for neutral Higgs at LEP 200}, \ccm{presented by Sau Lan Wu}, in \pt{{\it Proc. of the ECFA Workshop on LEP 200}}, Vol.~\vol{II}, \cp{Aachen, Germany}, eds. \efm{A.} \esm{Bohm} and \efm{W.} \esm{Hoogland} (\pn{CERN}, \y{1987}), p.~\pgs{312}.

\bibitem{ALEPH} \ccm{ALEPH, DELPHI, L3 and OPAL Collabs., CERN EP/2001-005}.

\bibitem{Namura} \afm{P. A.} \asm{McNamara} and \afm{S. L.} \asm{Wu}, \jat{Rep. Prog. Phys.} \vol{{\bf 65}}, \pgs{465} (\y{2002}).

\bibitem{CDFD03} \ccm{CDF and D0 Collabs., FERMILAB-CONF-12-318-E (2012)}.

\bibitem{CDFD04} \org{CDF and D0 Collabs.}, \jat{Phys. Rev. Lett.} \vol{{\bf 109}}, \pgs{071804} (\y{2012}).

\bibitem{Yellowbook} \ccm{CERN Yellow Report - Handbook of LHC Higgs cross sections: Books 1, 2 and 3. Report of the LHC Higgs Cross Section Working Group, Editors: S. Heinemeyer, C. Mariotti, G. Passarino and R. Tanaka, arXiv:1101.0593v3 [hep-ph] 20 May 2011, arXiv:1201.3084v1 [hep-ph] 15 Jan 2012, arXiv:1307.1347v1 [hep-ph] 4 Jul 2013. See also The Higgs Cross Section Working Group web page: \url{https://twiki.cern.ch/twiki/bin/view/LHCPhysics/CrossSections}, 2013 }.

\bibitem{ATLAS} \ccm{ATLAS Collab., CERN/LHCC/92-4 (1992); CMS Collab., CERN/LHCC/ 92-3 (1992)}.

\bibitem{EPS} \ccm{W. Murray, {\it Proc. Science}, {\it PoS} {\bf (EPS-HEP2011)031}, 2011 {\it Europhysics Conf. on High-Energy Physics}, EPS-HEP 2011, July~21--27, 2011, Grenoble, Rh\^{o}ne-Alpes, France (SISSA, 2012)}.

\bibitem{Lepton} \ccm{A. Nisati (for the ATLAS Collab.), {\it Proc. of Lepton--Photon 2011} (August 2011), eds. R.~M. Godbole and N.~K. Mondal, {\it PRAMANA} {\bf 79}, 541 (2012)}.

\bibitem{ATLASPUBFig8} \ccm{ATLAS Collab., Phys. Rev. Lett. {\bf 108}, 111803 (2012)}.

\bibitem{CMSPUBFig8} \ccm{CMS Collab., Phys. Lett. B {\bf 710} (2012) 403-425}.

\bibitem{ATLASPUBFig9} \ccm{ATLAS Collab., Phys. Lett. B {\bf 710} (2012) 383-402}.

\bibitem{CMSPUBFig9} \ccm{CMS Collab., Phys. Rev. Lett. {\bf 108}, 111804 (2012)}.

\bibitem{ATLASPUBFig10} \ccm{ATLAS Collab., Phys. Lett. B {\bf 710} (2012) 49-66}.

\bibitem{CMSPUBFig10} \ccm{CMS Collab., Phys. Lett. B {\bf 710} (2012) 26-48}.

\bibitem{scientific} \afm{M.} \asm{Riordan}, \afm{G.} \asm{Tonelli} and \afm{S. L.} \asm{Wu}, \jat{Sci. Am.} \vol{{\bf 307}}, \pgs{66} (\y{2012}).



\bibitem{ACPUBSpin} \ccm{ATLAS Collab., Phys. Lett. B {\bf 726}, 120-144 (2013)}; \ccm{CMS Collab., CERN preprint CERN-PH-EP-2013-220, submitted to Phys. Rev. D}; \ccm{CMS Collab., JHEP {\bf 1401} 096 (2014)}; \ccm{CMS Collab., CMS-PAS-HIG-13-016 (2013)}.

\bibitem{Gastman} \ccm{R. Gastman, S. L. Wu and T. T. Wu, CERN preprints CERN-PH-TH/2011-200 and CERN-PH-TH/2011-201}.

\bibitem{Rizzo} \afm{T. G.} \asm{Rizzo}, \jat{Phys. Rev. D} \vol{{\bf 22}}, \pgs{178} (\y{1980}).

\end{thebibliography}
\end{document}